\documentclass[twocolumn,showpacs,prb]{revtex4}

\usepackage[dvips]{graphicx}% Include figure files
\usepackage{dcolumn}% Align table columns on decimal point
\usepackage{bm}% bold math

\begin{document}

\title{\normalsize Insulating Phases Induced by Crossing of Partially Filled Landau Levels 
 in a Si Quantum Well}

\author{Tohru Okamoto,$^1$ Kohei Sasaki,$^1$ Kiyohiko Toyama,$^1$ Ryuichi Masutomi,$^1$
Kentarou Sawano,$^2$ and Yasuhiro Shiraki$^2$}
\affiliation{$^1$Department of Physics, University of Tokyo, 7-3-1, Hongo, Bunkyo-ku, Tokyo 113-0033, Japan\\
$^2$Research Center for Silicon Nano-Science, Advanced Research Laboratories, Tokyo City University,
8-15-1 Todoroki, Setagaya-ku, Tokyo 158-0082, Japan}

\date{3 March 2009}

\begin{abstract}
We study magnetotransport in a high mobility Si two-dimensional electron system
by {\it in situ} tilting of the sample relative to the magnetic field.
A pronounced dip in the longitudinal resistivity
is observed during the Landau-level crossing process
for noninteger filling factors.
Together with a Hall resistivity change which exhibits the particle-hole symmetry,
this indicates that electrons or holes in the relevant Landau levels become localized
at the coincidence where the pseudospin-unpolarized state is expected to be stable.

\end{abstract}

\maketitle

In a strong magnetic field, the single-particle energy spectrum 
of a two-dimensional electron system (2DES)
is quantized into Landau levels (LLs) 
and the spin degeneracy is lifted due to the Zeeman effect.
The ratio of the Zeeman energy $E_z$ to the cyclotron energy $\hbar \omega_c$ can be controlled
by tilting the magnetic field with respect to the sample normal,
where $\omega_c$ is the cyclotron frequency.
A crossing of LLs with different orbital and spin indices occurs
when $E_z$ becomes comparable to $\hbar \omega_c$.
In the vicinity of the coincidence,
 where the two LLs are nearly degenerate in the single-particle energy,
the electron-electron interaction plays an essential role in determining the ground-state configuration.
To study many-body states and quantum phase transitions,
it is convenient to relabel the two crossing LLs
in terms of pseudospin.\cite{MacDonald1990,Jungwirth1998,Jungwirth2000}
The $z$ component of the pseudospin magnetization unit vector can be defined
as $m_z=+1$ ($-1$) when all the relevant electrons reside
in the first (second) LL.
For integer LL filling factors $\nu$,
it is predicted that the ground state is the ferromagnetic state with easy-axis (Ising) anisotropy.
\cite{Jungwirth1998,Jungwirth2000}
The magnetization jumps discontinuously at a point where the $m_z=+1$ and $m_z=-1$ states are degenerate,
i.e., the zero point of the pseudospin field.
Experimentally, resistance spikes and hysteresis have been observed for several 2DESs 
(Refs. \onlinecite{Piazza1999,Poortere2000,Jaroszynski2002,Poortere2003,Chokomakoua2004,Toyama2008})
and discussed in association with domain wall between $m_z=+1$ and $m_z=-1$ regions of the sample. \cite{Jungwirth2001}
On the other hand, the pseudospin configuration has not been studied 
for the intermediate region between integer QH states.
In the case of real spin systems, 
the ground state was found to be spin unpolarized
for various noninteger values of $\nu$ when the bare Zeeman splitting is sufficiently small.
\cite{Eisenstein1988,Clark1989,Eisenstein1989,Engel1992,Barrett1995,Du1995,Freytag2002}
It is likely that an unpolarized ground state would also occur
in pseudospin systems for $\nu \neq$~integer
and small pseudospin field.
The pseudospin configuration with $|m_z|<1$ corresponds to a new quantum state
characterized by an admixture of electrons with different orbital wave functions.

In this paper, we report magnetotransport measurements on a high-mobility Si 2DES
in the vicinity of the coincidence of LLs.
{\it In situ} tilting of the sample relative to the magnetic field allows 
independent control of $\nu$ and $E_z/\hbar \omega_c$.
In the intermediate region, we observe a pronounced dip in the longitudinal resistivity
and a Hall resistivity change which exhibits the particle-hole symmetry.
The results indicate that electrons or holes in the relevant Landau levels
become localized at the coincidence where the pseudospin is expected to be unpolarized.
The $\nu$ dependence of the dip width implies that
the pseudospin-unpolarized phase develops as $\nu$ decreases 
from an integer.
Hysteresis behavior that occurs around the dip is interpreted as evidence for the first-order transition
between the pseudospin-unpolarized and pseudospin-polarized states.

We used a Si/SiGe heterostructure with a 20-nm-thick strained Si channel sandwiched between relaxed $\mathrm{Si}_{0.8}\mathrm{Ge}_{0.2}$ layers. \cite{Yutani1996}
The electrons are provided by a Sb-$\delta$-doped layer 20~nm above the channel.
The electron concentration $N_s$ can be controlled by varying bias voltage $V_{\rm BG}$
 of a $p$-type Si(001) substrate 2.1~$\mu$m below the channel at 20~K.
Electron spin resonance measurements have shown that the spin-orbit interactions and the electron-nuclear spin (hyperfine) coupling are very small in the present 2DES. \cite{Matsunami2006}
Two Hall bars, denoted by HB-X and HB-Y, were fabricated on the same chip as illustrated in Fig.~1(a)
to study the anisotropy with respect to the in-plane magnetic field direction. \cite{Toyama2008,Zeitler2001}
\begin{figure}[t!]
\includegraphics[width=7cm]{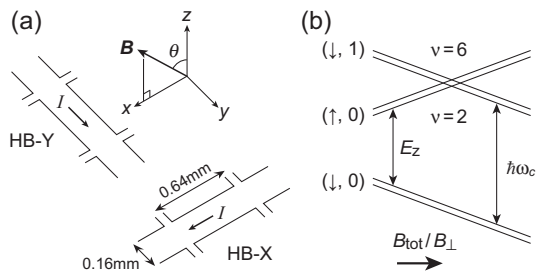}
\caption{
(a) Schematic of the Hall bars.
The current direction is along the $x$ axis in HB-X and the $y$ axis in HB-Y.
The magnetic field is tilted from the $z$ axis in the $x$ direction by an angle $\theta$.
(b) Landau levels in a tilted magnetic field.
The ratio of the Zeeman energy $E_z$ to the cyclotron energy $\hbar \omega_c$ increases with $\theta$.
$E_z$ and $\hbar \omega_c$ are approximately proportional to 
the total magnetic field $B_{\rm tot}$
and the perpendicular component $B_\perp = B_{\rm tot} \cos \theta$, respectively.
}
\end{figure}
HB-X (HB-Y) is oriented along the $x$ direction ($y$ direction) and used for the measurements of
the longitudinal resistivity $\rho_{xx}$ ($\rho_{yy}$) and
the Hall resistivity $\rho_{yx}$ ($\rho_{xy}$).
After brief illumination of a red light emitting diode with $V_{\rm BG}=-4.2$~V,
$N_s$ and low-temperature mobility $\mu$ are 
1.38 (1.43) $\times 10^{11}~\mathrm{cm^{-2}}$
and 2.6 (2.8) $\times 10^{5}~\mathrm{cm^2/V~s}$, respectively,
for HB-X (HB-Y). \cite{HigherNs}
The sample was mounted on a rotatory stage in a ${}^3$He-${}^4$He dilution refrigerator
and the in-plane magnetic field was oriented along the $x$ direction.
To avoid the heating effect, a small ac current of 1.0~nA was used in a standard low-frequency (11.7~Hz) lock-in technique.

Here we focus on the coincidence of the spin-up LL with an orbital index of $n=0$ [LL$(\uparrow, 0)$] and the spin-down LL with $n=1$ [LL$(\downarrow, 1)$].
We assign $m_z=+1$ to LL$(\uparrow,0)$ and $m_z=-1$ to LL$(\downarrow,1)$.
The Fermi level $\varepsilon_F$ is inside these degenerate LLs when the Landau-level filling factor $\nu$
is in the range between 2 and 6 as shown in Fig.~1(b).
Note that we have a two-fold valley degeneracy for a 2DES formed on the Si(001) surface. \cite{Ando1982}
In our experimental setup, the tilt angle $\theta$ can be continuously varied 
while the temperature $T$ is kept constant at a fixed value.
Furthermore, simultaneous control of $\theta$ and the total magnetic field $B_{\rm tot}$
allows us to measure the $B_{\rm tot}$-dependence
for a constant value of the perpendicular component $B_\perp=B_{\rm tot} \cos\theta$,
which is proportional to $\nu$.

Figure 2 shows $\rho_{xx}$ and $\rho_{yx}$ as functions of $B_\perp$
in the range between the QH states of $\rho_{yx}=h/3e^2$ and $h/2e^2$.
\begin{figure}[t!]
\includegraphics[width=7cm]{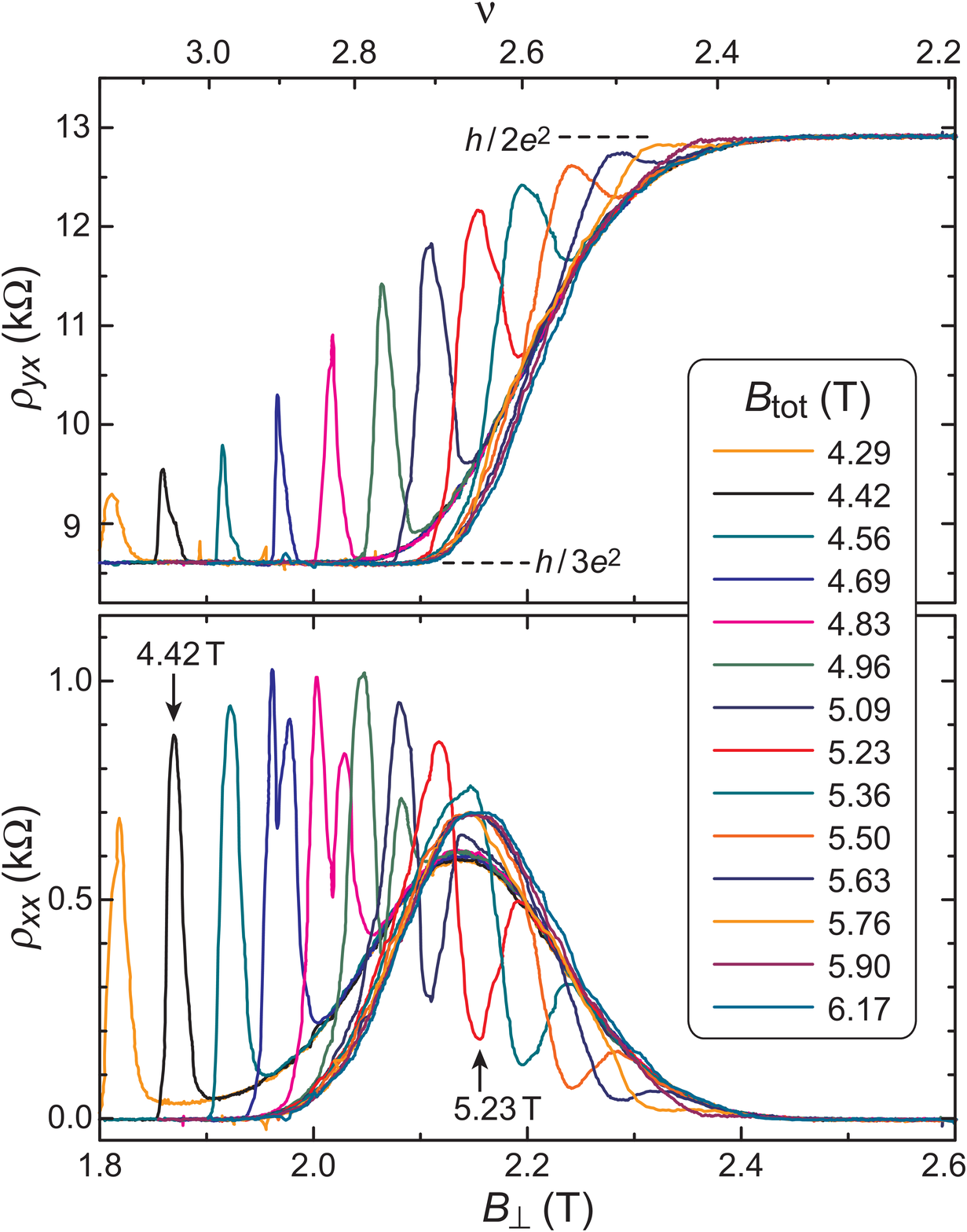}
\caption{(Color online)
$B_\perp$ dependence of $\rho_{xx}$ and $\rho_{yx}$ 
for different fixed values of $B_{\rm tot}$.
The data were obtained for HB-X by decreasing $B_\perp$ at 70~mK.
The position of the peak or dip shifts to higher $B_\perp$ as $B_{\rm tot}$ increases.
The peak for $B_{\rm tot} =4.42$~T and the dip for $B_{\rm tot} = 5.23$~T are indicated by arrows.
}
\end{figure}
Data for different fixed values of $B_{\rm tot}$ are presented.
An anomalous structure arising from the LL crossing is clearly seen in each curve
and the position shifts to higher $B_\perp$ as $B_{\rm tot}$ increases.
In the QH region near $\nu=3$ (e.g., $B_\perp \approx 1.87$~T and $B_{\rm tot} = 4.42$~T), 
$\rho_{xx}$ exhibits a sharp spike
which is attributed to domain wall resistance in Ising QH ferromagnets.
\cite{Poortere2000,Jaroszynski2002,Poortere2003,Chokomakoua2004,Toyama2008}
On the other hand, a pronounced dip is observed in the intermediate region
(e.g., $B_\perp \approx 2.15$~T and $B_{\rm tot} = 5.23$~T).
We believe that the dip has a different origin than the spike
and is related to the occurrence of the pseudospin-unpolarized phase discussed later.
It appears even in the spike
and leads to the apparent double-peak structure around $B_\perp \approx 2.0$~T.
While the structure of the longitudinal resistivity
at the coincidence is sensitive to $\nu$,
the change in $\rho_{yx}$ is always positive in this range.
The position of the peak center is in good agreement with
the center of the $\rho_{xx}$ dip in the intermediate region.

As reported in previous studies on Si/SiGe heterostructures in the vicinity of $\nu = 4$, \cite{Toyama2008,Zeitler2001}
the resistivity spike in the QH region exhibits a strong anisotropy with respect to the in-plane magnetic field.
In Figs. 3(a) and 3(b), results for $\nu \approx 3$ are shown. \cite{Crystallographic}
\begin{figure}[t!]
\includegraphics[width=7cm]{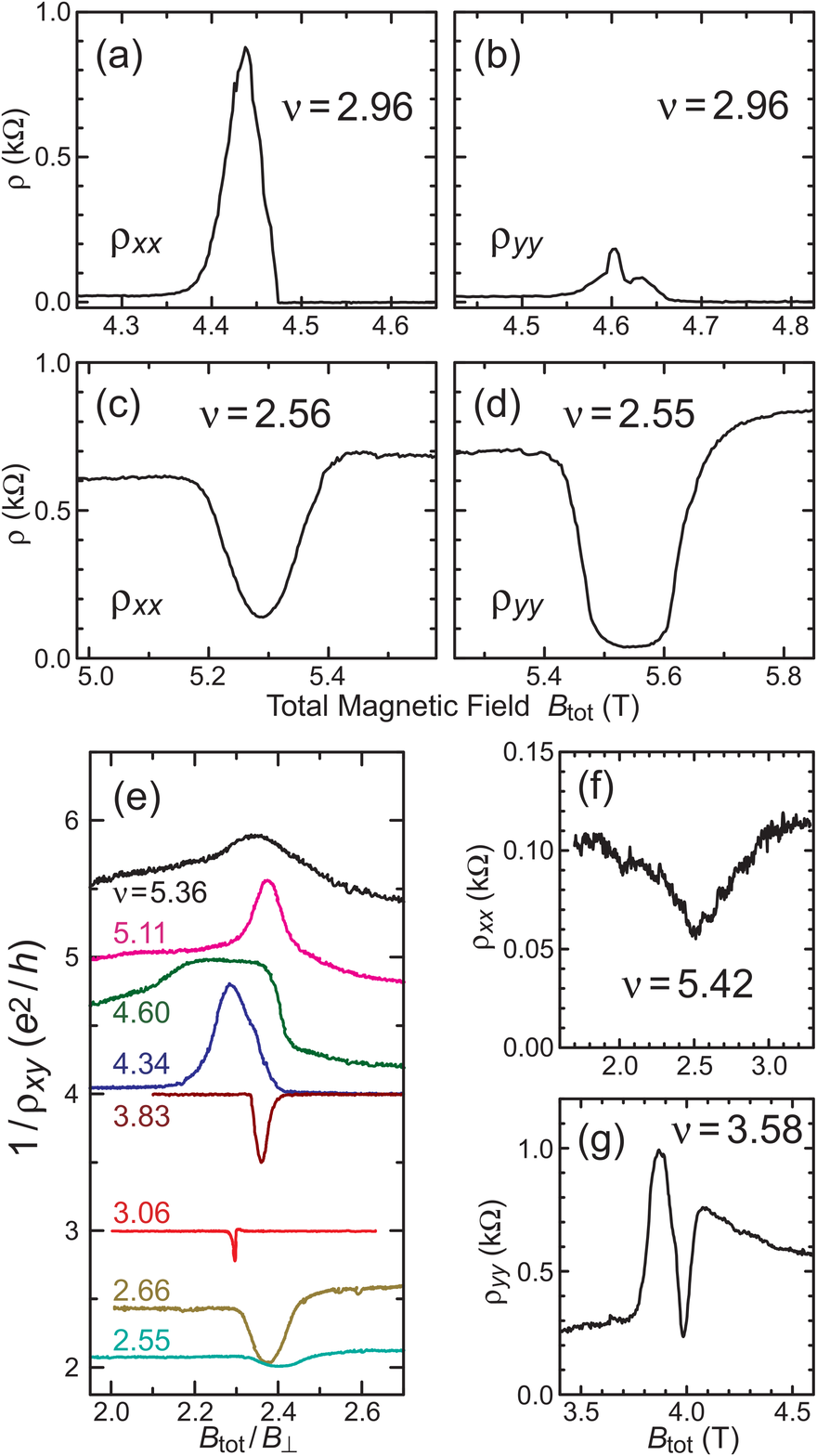}
\caption{(Color online)
[(a)-(d), (f), (g)] Longitudinal resistivities as functions of $B_{\rm tot}$
for different fixed values of $\nu$.
(e) Inverse of the Hall resistivity vs $B_{\rm tot}/B_\perp$.
All the data were obtained by decreasing $B_{\rm tot}$ at 70~mK.
HB-X was used for (a), (c), and (f).
HB-Y was used for (b), (d), (e), and (g).
}
\end{figure}
The anisotropy has been discussed in terms of a stripe-like domain structure 
of the Ising QH ferromagnet oriented perpendicular to the in-plane magnetic field. \cite{Toyama2008,Zeitler2001}
However, the dip in the intermediate region does not show a well-defined anisotropy.
Typical data set obtained for the range of $2<\nu<3$ is presented in Figs. 3(c) and 3(d).
This also supports that the origin of the dip is different than that of the spike in the Ising QH ferromagnet.

In the range of $2<\nu<6$ where $\varepsilon_F$ is inside the four degenerate LLs,
particle-hole symmetry centered at $\nu=4$ is expected to occur.
It appears clearly in a Hall resistivity change.
In Fig.~3(e), the inverse of $\rho_{xy}$ in the vicinity of the LL coincidence is shown.
The sign of the change in $\rho_{xy}$ is positive for $\nu <4$
and negative for $\nu > 4$.
Similar behavior was observed for $\rho_{yx}$.
This indicates that the number of mobile electrons or holes that carry the Hall current
is reduced at the LL coincidence.
As shown in Fig.~3(f), the longitudinal resistivity exhibits a dip also for $5<\nu<6$
where the hole picture is useful. \cite{eh}
The dip in the longitudinal resistivity accompanied with the Hall resistivity change
reminds us of insulating phases in spin-resolved high LLs of high-mobility GaAs 2DESs, \cite{Lilly1999,Du1999,Cooper1999}
which are interpreted as bubble phases of electrons or holes. \cite{Koulakov1996,Fogler1996,Moessner1996,Haldane2000,Shibata2001,Lewis2002}
Although the mechanism leading to the insulating state is unclear for our system,
the results strongly suggest that
electrons or holes become localized at the LL coincidence 
at least for $2<\nu<3$ and $5<\nu<6$.

In the ranges of $3<\nu<4$ and $4<\nu<5$,
peculiar behaviors are observed possibly due to the valley degeneracy
[e.g., Fig.~3(g)].
As discussed in our previous work for $\nu=4$, \cite{Toyama2008}
the bare valley splitting is estimated to be very small in the present 2DESs.
At $\nu=4$, the LL crossing is observed as a single narrow spike in $\rho_{xx}$,
indicating that the valley degeneracy is not lifted.
On the other hand, the many-body effect strongly enhances the valley splitting
and stabilizes the QH state at $\nu={\rm odd}$.
The valley polarization in the intermediate region between $\nu=4$ and $\nu=3$ (or 5)
is unclear and might be altered during the LL crossing process.
In fact, the shape of the resistivity change, which is sensitive to $\nu$, is not simple in these ranges.
In the following, we shall not further discuss the valley state,
and restrict ourselves to the range of $2<\nu<3$.

In Fig.~4(a), the width $\Delta B_{\rm tot}$ of the insulating state determined from 
the dip in the longitudinal resistivities is shown.
\begin{figure}[t!]
\includegraphics[width=7cm]{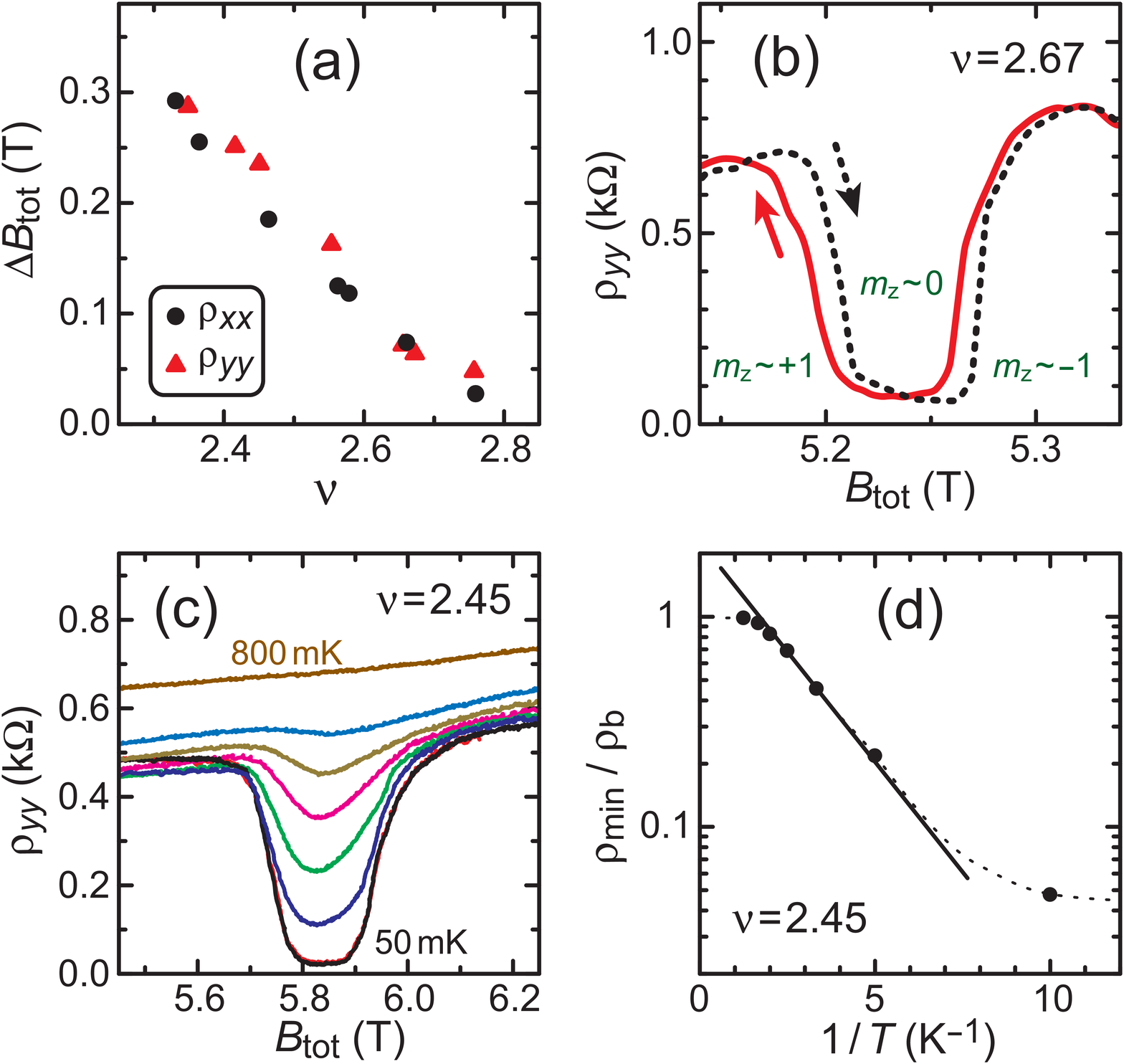}
\caption{(Color online)
(a) The full width at half maximum of the dip in $\rho_{xx}$ or $\rho_{yy}$
measured at 70~mK.
Since hysteresis was observed, the mean values are plotted. 
(b) Hysteresis of $B_{\rm tot}$-dependence of $\rho_{yy}$ at 70~mK and $\nu=2.67$ ($B_\perp=2.21~{\rm T}$).
Solid curve (red): the downward sweep with $dB_{\rm tot}/dt=-0.5$~mT/s.
Dashed curve (black): the upward sweep with $dB_{\rm tot}/dt=+0.7$~mT/s.
(c) $B_{\rm tot}$-dependence of $\rho_{yy}$ at $\nu=2.45$ ($B_\perp=2.41~{\rm T}$)
for $T=50$, 100, 200, 300, 400, 500, 600, and 800~mK (from bottom to top).
The data were obtained by decreasing $B_{\rm tot}$.
(d) Logarithm plot of the $\rho_{yy}$ minimum, divided by the baseline value $\rho_b$, is shown 
as a function of the inverse temperature.
The solid straight line represents
an Arrhenius temperature dependence $\rho_{yy}/\rho_b \propto \exp(-\Delta /2T)$,
where $\Delta =1.0$~K is the energy gap.
The dotted line is a guide for the eyes.
}
\end{figure}
It decreases almost linearly with increasing $\nu$
and goes to zero.
At $\nu=3$, the Ising QH ferromagnetic state with $m_z=+1$ or $m_z=-1$ is expected to be stable
even for zero pseudospin field.
The $\nu$ dependence of $\Delta B_{\rm tot}$ implies that
the pseudospin-unpolarized phase with $|m_z| <1$ develops as $\nu$ decreases in the range of $2< \nu <3$. 
As shown in Fig.~4(b), the transition between conducting and insulating states
exhibits hysteresis, which is a common feature of first-order transitions.
Since the hysteresis loops for high-$B_{\rm tot}$ and low-$B_{\rm tot}$ transitions are
similar to each other,
it is expected that the jumps in $m_z$ are also comparable.
This means that the pseudospin magnetization is small in the insulating region.
The dip in the longitudinal resistivity disappears
as $T$ is raised.
The data for $\rho_{yy}$ at $\nu=2.45$ are shown in Fig.~4(c).
If we assume an Arrhenius temperature dependence, 
the energy gap is deduced to be $\Delta \sim 1$~K as exemplified in Fig.~4(d).

The state with small $m_z$ corresponds to a configuration in which
the LL$(\uparrow,0)$ and LL$(\downarrow,1)$ are nearly equally populated.
Akera, MacDonald, and Yoshioka have studied the ground state of
a double-layer system in which a higher LL ($n=4$ or 5)
in one layer is degenerate with the ground LL ($n=0$) in the other. \cite{Akera1995}
It was shown that the ground state can be pseudospin unpolarized
for small values of the effective filling factor $\nu^\ast=\nu-n$
in the limit of zero layer separation,
where the system is equivalent to a single-layer system
with the spin degree of freedom and nonzero Zeeman splitting.
By an approach using the analytical model wave function,
the pseudospin-unpolarized ground state was found to be an electron crystal, 
which is composed of two interpenetrating square sublattices.
In order to study the ground-state pseudospin polarization in our system,
we extend the calculations in Ref.~\onlinecite{Akera1995}
to the case of $n=1$ for the higher LL.
Although effects of layer thickness, LL mixing, and the valley degree of freedom
might be important, we do not consider them for simplicity.
Exact diagonalizations on finite clusters with four electrons
have shown that the pseudospin-unpolarized phase occurs
at least for $\nu^\ast \lesssim 0.3$
while the results are not conclusive for higher $\nu^\ast$.

In summary, we have performed the magnetotransport measurements on a high-mobility Si 2DES
in the vicinity of the coincidence of LL$(\uparrow, 0)$ and LL$(\downarrow, 1)$.
In contrast to the QH region where $\rho_{xx}$ shows a narrow spike,
we found a pronounced dip in the intermediate region.
The sign of the Hall resistivity change is positive for $\nu<4$
and negative for $\nu>4$.
These results indicate that electrons or holes in the relevant Landau levels
become localized at the coincidence where the pseudospin is expected to be unpolarized.
The $\nu$ dependence of $\Delta B_{\rm tot}$ implies that
the pseudospin-unpolarized phase develops as $\nu$ decreases in the range of $2<\nu <3$. 
Hysteresis loops observed around the dip were interpreted as evidence for the first order transition
between the pseudospin-unpolarized and pseudospin-polarized states.
The energy gap was deduced to be $\Delta \sim 1$~K from the Arrhenius plot of the longitudinal resistivity.

This work has partly supported by Grant-in-Aids for Scientific Research (B) (Grant No. 18340080), (A) (Grant No. 21244047) and Grant-in-Aid for Scientific Research on Priority Area ``Physics of New Quantum Phases in Superclean Materials'' (Grant No. 20029005) from MEXT, Japan.


\begin{thebibliography}{30}
\bibitem{HigherNs}
Qualitatively similar experimental results were obtained for $V_{\rm BG}=0$~V
with $N_s=2.3 \times 10^{11}~\mathrm{cm^{-2}}$
and $\mu=6\times 10^{5}~\mathrm{cm^2/V~s}$.
\bibitem{Crystallographic}
We have also performed measurements in a magnetic field 
where the in-plane component is aligned parallel to the $y$ direction.
It was confirmed that the transport anisotropy in the QH region
is dominated by the direction of the in-plane magnetic field with respect
to the current channel, not by the crystallographic anisotropy.
\end{thebibliography}
\end{document}